\documentstyle{ichep_preprint}
\input{psfig}

\def\ie{{\it i.e.}}
\def\epem{e^+e^-}
\def\gsim{{\buildrel{>}\over{\sim}}}
\def\lsim{{\buildrel{<}\over{\sim}}}
\def\gluino{\widetilde g}
\def\chitil{\widetilde\chi}
\def\mgluino{m_{\widetilde g}}
\def\mt{m_t}

\def\mgut{M_{U}}
\def\rbtau{R_{b/\tau}}

\def\sms{SMS}
\def\ms{MS}
\def\dil{D}
\def\sdil{SD}
\def\msp{\ms^+}
\def\msm{\ms^-}
\def\dilp{\dil^+}
\def\dilm{\dil^-}
\def\smsp{\sms^+}
\def\smsm{\sms^-}
\def\sdilp{\sdil^+}
\def\sdilm{\sdil^-}

\def\mgltb{$\mgl\,$--$\tanb$}
\def\hl{h^0}
\def\hh{H^0}
\def\ha{A^0}
\def\hm{H^-}
\def\hp{H^+}
\def\mhl{m_{\hl}}

\def\hpm{H^{\pm}}
\def\mhpm{m_{\hpm}}
\def\hp{H^+}
\def\gam{\gamma}

\def\gev{~{\rm GeV}}
\def\tev{~{\rm TeV}}
\def\pbi{~{\rm pb}^{-1}}
\def\fbi{~{\rm fb}^{-1}}
\def\fb{~{\rm fb}}

\def\mt{m_t}

\def\rta{\rightarrow}

\def\anti{\overline}
\def\pbi{~{\rm pb}^{-1}}
\def\fbi{~{\rm fb}^{-1}}
\def\fb{~{\rm fb}}

\catcode`\@=11 
\def\Tr{\mathop{\rm Tr}\nolimits}

\def\t1{{\tilde 1}}

\def\slash#1{#1\hskip-10pt/\hskip10pt}

\def\etmiss{\slash {E_T}}

\def\gev{\,{\rm GeV}}
\def\tev{\,{\rm TeV}}

\def\wt{\widetilde}

\def\rta{\rightarrow}

\def\gl{\wt g}
\def\mgl{m_{\gl}}

\def\stopone{\wt t_1}

\def\mstopone{m_{\stopone}}
\def\sq{\wt q}

\def\slep{\wt l}

\def\slepl{\slep_L}
\def\slepr{\slep_R}
\def\mslepl{m_{\slepl}}
\def\mslepr{m_{\slepr}}

\def\hl{h^0}
\def\hh{H^0}
\def\ha{A^0}
\def\hp{H^+}
\def\hm{H^-}
\def\hpm{H^{\pm}}
\def\mhl{m_{\hl}}

\def\mhp{m_{\hp}}
\def\tanb{\tan\beta}
\def\mt{m_t}
\def\mb{m_b}
\def\mz{m_Z}
\def\mgut{M_U}
\def\mstring{M_S}

\def\cnone{\wt\chi^0_1}
\def\cntwo{\wt\chi^0_2}
\def\snu{\wt\nu}

\def\msnu{m_{\snu}}
\def\mcnone{m_{\cnone}}
\def\mcntwo{m_{\cntwo}}

\def\cpone{\wt \chi^+_1}
\def\cmone{\wt \chi^-_1}
\def\cpmone{\wt \chi^{\pm}_1}
\def\mcpone{m_{\cpone}}
\def\mcpmone{m_{\cpmone}}
\def\stau{\wt\tau}

\def\tanb{\tan\beta}

\def\mw{m_W}
\def\mz{m_Z}
\def\anti{\overline}

\def\ifmath#1{\relax\ifmmode #1\else $#1$\fi}

\def\3quarter{{\textstyle{3 \over 4}}}

\begin{document}

\title{Supersymmetry at LEPII, the Tevatron, and Future Accelerators
*
}

\author{J. F. Gunion$^{\dag\ddag}$}

\affil{\dag\ Davis Institute for High Energy Physics,\\
Department of Physics, University of California at Davis, Davis CA
95616, U.S.A.}

\abstract{I present a brief review of prospects for discovering
supersymmetry, focusing primarily on the Tevatron and LEPII, but
commenting on the LHC and a next linear $\epem$ collider.
Special emphasis is given
to expectations in the context of models with GUT boundary
conditions motivated by Supergravity and Superstrings.
An overview of related conference contributions is given.
\cabs          
}

\twocolumn[\maketitle]

\fnm{7}{To appear in the Proceedings of the 27th International Conference
on High Energy Physics, Glasgow, July 1994, eds. P. Bussey and I. Knowles.}
\fnm{2}{Work supported, in part, by U.S. DOE.}

\section{Introduction}
The success of gauge coupling unification in the context of the Minimal
Supersymmetric Model (MSSM) lends considerable credence not only to the
possibility that this extension of the Standard Model
is correct, but also to the idea that the boundary conditions
for all the soft-supersymmetry-breaking parameters at
the unification scale could be relatively simple and universal.
Supergravity and superstring theory each provide particularly
attractive and well-motivated examples of such boundary conditions.
These, and their phenomenological implications, especially
for the Tevatron and LEPII,
are briefly reviewed, with emphasis on the very real possibility that
supersymmetry could be discovered at these two (highly
complementary) accelerators. My discussion will be based
on the study of Ref.~\cite{bgkp}. References to the many related
studies are given therein; in particular, see Ref.~\cite{lopez}.

The four basic parameters of supersymmetry
breaking are: a) the gaugino masses $M_a$ (where $a$ labels the
group); b) the scalar masses $m_i$ (where $i$ labels the various
scalars, e.g. Higgs, sleptons, squarks); c) the soft Yukawa
coefficients $A_{ijk}$; and d) the $B$ parameter which specifies the
soft mixing term between the two Higgs scalar fields. The
supersymmetry-breaking schemes considered here will be: i)
`no-scale' or minimal supergravity (labelled by MS) and ii) dilaton-like
superstring (labelled by D). Predictions in these models for the $B$ parameter
are rather uncertain, and so we shall leave it a free parameter. In the MS
models the only other source of supersymmetry breaking is via the
gaugino masses $M_a$, which are taken to have a universal value $M_a\equiv
M^0$ at $\mgut$; the $m_i^0$ and $A_{ijk}^0$ are taken to be zero. In the D
models the $M_a$, $m_i$ and $A_{ijk}$ parameters all take on universal values
at $\mgut$ related by:
\begin{equation}
M^0=-A^0=\sqrt3m^0.
\end{equation}
These latter dilaton-like boundary conditions are certainly those appropriate
when supersymmetry breaking is dominated by the dilaton field in
string theory, but also apply for a remarkably broad class of models
(including Calabi-Yau compactifications, and orbifold models in which
the MSSM fields all belong to the untwisted sector) so long
as the moduli fields do not play a dominant role in supersymmetry breaking.
For a brief review and detailed references, see Ref.~\cite{bgkp}.

Given either choice of boundary conditions, if the top quark mass is fixed (we
shall quote results for $\mt(\mt)=170\gev$, corresponding to a pole mass of
about 178 GeV) only two free parameters and a sign remain undetermined
after minimizing the potential. The two parameters can be taken to be
$\tanb$, the ratio of the neutral Higgs field vacuum expectation values, and
$\mgluino$, the gluino mass. The parameter $B$ is determined in terms of
these, as are all other superpotential parameters, including the magnitude of
the Higgs superfield mixing parameter $\mu$. However, the sign of $\mu$ is
not determined. Four models result --- $\msp$, $\msm$, $\dilp$,
$\dilm$, the superscript indicating the sign of $\mu$ --- the phenomenology
of which can be explored in the two dimensional \mgltb\ parameter space.

The discussion so far has obscured one fundamental problem facing
the gauge coupling unification success: namely, the scale  $\mgut$
at which the couplings naturally unify is $\sim 2\times 10^{16}\gev$,
i.e. much less than the natural scale for supergravity and string unification
of
$\mstring\sim10^{18}\gev$.  A variety of excuses for this have been
discussed. In Ref.~\cite{bgkp} two extreme approaches are adopted:
i) ignore the difference ---
a fuller understanding of the feed-down of SUSY breaking from the
full supergravity or superstring theory could resolve the discrepancy;
ii) assume that the unification at $\mgut$ is only apparent (i.e. accidental)
and introduce a minimal set of additional matter fields at high scale
with masses chosen precisely so as to give coupling unification
at $\mstring$.  I will not go into details regarding these extra fields;
a discussion and references can be found in Ref.~\cite{bgkp}.
The models with such extra fields will be termed the `string-scale-unified'
versions of the previously listed models, and will be denoted by
$\smsp$, $\smsm$, $\sdilp$, and $\sdilm$.

\vspace{.5cm}
\begin{figure}[htbp]
\begin{center}
\leavevmode
{\psfig{figure=ichep_fig.ps,width=3in,clip=}}
\end{center}
{\caption{
Parameter space boundaries and discovery contour
limits for the $\dil$ and $\sdil$ models.
LEPII discovery limits for $\slepr\slepr$,
$\cpone\cmone$, $Z\hl$ are the dotted, short-dash, and
dot-dash curves, respectively; Tevatron `SS plus
$3\ell$' discovery limits are the long-dash curves.
 }}
\label{dilfig}
\end{figure}

\section{Phenomenology}

To systematically investigate the resulting 8 models, we first establish the
allowed region of \mgltb\ parameter space for each subject to: a) all
predicted SUSY partner particles (including the light Higgs $\hl$) are
unobservable; b) the lightest SUSY particle is either the lightest neutralino
$\cnone$ (as is always the case for the allowed parameter space
of the models explored here) or the
sneutrino $\snu$; c) the top quark Yukawa remains perturbative at all scales
from $\mw$ up to $\mgut$ or $\mstring$; d) proper electroweak symmetry
breaking and a global minimum are obtained. We do not impose $b\rta s\gamma$,
relic abundance, or proton decay constraints, as these all have
considerable uncertainties and/or require additional model-dependent input.
We also do not require exact $b-\tau$ Yukawa unification. \Fref{dilfig}
shows the $\dil$ and $\sdil$ parameter space boundaries; $\sms$ results
are similar.
The lower $\mgluino$ limit is set either by $\mslepr<\mz/2$ or $\mcpone<\mz/2$.
The upper limit on $\tanb$ results from requiring that the lighter $\stau$
eigenstate not be the LSP.
The lower limit on $\tanb$ results from the requirement that $\lambda_t$
remain perturbative at all RGE scales. In the $\dil$, $\sdil$, and $\sms$
models there is no upper limit on $\mgluino$ (but $\mgluino\lsim 700\gev$
for $\ms$ models),
although standard considerations of naturalness and accurate gauge coupling
unification both suggest that $\mgluino\lsim 1\tev$. Note that
the deviations from exact $b-\tau$ unification,
$\rbtau\equiv\lambda_b/\lambda_\tau=1$ (see the right-hand axis
of the figure), are at most 25\%.

Within the allowed parameter spaces shown, the masses of the
SUSY particles scale with $\mgluino$; variation of the masses with $\tanb$
at fixed $\mgluino$ is relatively limited for $\mgluino$
values above about 500 to 600 GeV, where one finds
$\mcnone<\mslepr,\mcpone,\mcntwo<\msnu<\mslepl$,
with $\slepr,\cpone,\cntwo,\snu,\slepl$ clustering between 0.2 to 0.4 times
$\mgluino$. First and second generation squark masses are of order
$\mgluino$, while the lighter stop is substantially lighter, roughly
$\mstopone\sim 0.7\mgluino$. It is the restricted size of the soft scalar
mass parameter, $m^0$, relative to $M^0$ that causes the sleptons to be
rather light in both the minimal-supergravity and dilaton-like models.
Indeed, slepton masses are largely generated by renormalization-group
evolution from the $M^0$ gaugino seed value at $\mgut$; only the squarks
acquire masses comparable to $\mgluino$, as a result of the driving terms
proportional to $\alpha_s$ in the RGE's.

Despite these broad similarities of all the models, there are important
details that vary as a function of model, $\tanb$ and $\mgluino$.
The most important issue for Tevatron phenomenology is whether
$\cntwo\rta \nu\snu$ and $\cpone\rta \ell^+\snu$ are kinematically
allowed or not.  When allowed, the $\snu$ decays invisibly, $\cntwo$
decays are dominated by the {\it invisible} $\nu\snu$ channel (especially
for $\mu<0$), and the $\ell$ in $\cpone$ decays is relatively soft.
When the $\cntwo$ decays invisibly and the $\ell$
from $\cpone$ is soft, SUSY detection at the Tevatron becomes
difficult.  Note that it is the low scalar slepton and sneutrino
masses predicted in the $m^0<M^0$ models considered here that allow
a delicate variation as to which channels are allowed.

LEPII is less sensitive to this kind of detail since it will generally
discover anything for which adequate energy is available.
The fairly similar mass scales for sleptons and charginos as a
function of \mgltb\ in the various models imply similar
discovery limits for the models at LEPII. To specify the actual discovery
boundaries in parameter space we assume $\sqrt s=200\gev$,
with integrated luminosity of $L=500\pbi$. In this case one
can observe $\slepr\slepr$, $\cpone\cmone$ pair production for
$\mslepr,\mcpone<95\gev$.  The resulting discovery limits are indicated by
the dotted and short-dashed lines in \Fref{dilfig}. Note the large
equivalent $\mgluino$ values probed, $\sim 300-400\gev$. Further,
$Z\hl$ production can be observed for $\mhl\lsim
105\gev$, i.e. the portion of parameter space to the left of the
dot-dashed line.

To explore the Tevatron discovery limit, we employed ISASUSY \cite{isasusy}
to generate all types of SUSY production processes, including complete
decay chains.  Events were generated for a series of \mgltb\ choices
within the allowed parameter domain for each model.  At Tevatron energies
$\chitil\chitil$ production,
including especially $\cntwo\cpmone$, has the largest rate.
Next in importance are
$\slep\slep,\slep\snu,\snu\snu$, $\gluino\chitil$ and $\sq\chitil$, as well
as $\stopone\stopone$, with $\gluino\gluino,\gluino\sq,\sq\sq$ processes
being relatively small (due to the large $\gluino$ and $\sq$ masses
compared to the Tevatron energy).

We then imposed cuts appropriate to looking for a variety of
different types of signals.  After examining backgrounds for these same cuts,
we found that for $L=1\fb$ only three types of signal were useful: i) the
missing energy ($\etmiss$) signal; ii) the same-sign (SS) dilepton
signal; and iii) the
tri-lepton ($3\ell$) signal.   More or less independent of model, the
$\etmiss$ signal will allow SUSY detection out to about $\mgluino=300\gev$.
The missing energy signal has significant contributions from
many types of production processes, including, in particular, $\cntwo\cpmone$
(where one of the leptons is missed or soft) and $\gluino\chitil,\sq\chitil$
--- $\gluino\gluino,\sq\gluino,\sq\sq$ sources contribute at a
lesser rate. The discovery boundaries arising from combining the SS and
$3\ell$ signals (both of which are essentially background-free for our cuts
--- we require 5 or more events in one or the other) are shown by the
long-dash curves in \Fref{dilfig}. Only in the $\dilm$ case is the
coverage of parameter space as limited
($\mgluino\lsim 300\gev$) as for the $\etmiss$ signal; in
the $\dilp$, $\sdilp$ and $\sdilm$ cases the $SS+3\ell$ signal
probes out to $\mgluino\lsim 500\gev$. As discussed earlier, this
difference is due to the dominance of the invisible
$\nu\snu$ decay mode of the the $\cntwo$ for the $\dilm$ case.
In the $\dilp$, $\sdilp$ and $\sdilm$ cases, the $\snu$ is sufficiently heavy
that $\cntwo\rta \nu\snu$ decays are kinematically forbidden, implying
significant 3-body decays to leptons for the $\cntwo$ and $\cpmone$.

Experimentally, in models with light sleptons the signature for $\etmiss$, SS,
or $3\ell$ events from $\cntwo\cpmone$ production is a small number of
associated jets. Indeed, events with few or no jets is the
most characteristic feature of models in which the $m^0$ parameter is smaller
than $M^0$. If sleptons and sneutrinos are heavy due to a large value for
$m^0$, then the final states resulting from the crucial $\cntwo\cpmone$
process will contain three leptons only part of the time. The smaller
number of purely leptonic final state events means that the SS and
$3\ell$ signals generally cannot probe much beyond $\mgluino\sim 300\gev$.
For a review of such scenarios see Ref.~\cite{baeretal}.

Finally, we note that \Fref{dilfig} delineates fairly clearly the role of the
LHC and a next linear $\epem$ collider (NLC)
for exploration of supersymmetry with minimal-supergravity or
dilaton-like boundary conditions. For gluino masses below $\mgluino\sim
500\gev$, discovery of supersymmetry at LEPII and/or the Tevatron would
be quite likely, but event rates could be very low.  The
gluino and heavier squarks would be likely to escape
detection unless $\mgluino$ is substantially below 500 GeV.  At the LHC
the $\gluino$ and $\sq$'s would be produced at high rates, allowing detailed
studies of their properties, including cascade decays.
An NLC with $\sqrt s\geq 500$ GeV would
be an extremely valuable complement, allowing precise determination
of the masses and decays of the inos appearing in the
$\gluino,\sq$ cascade decays.
For $\mgluino$ much beyond 500 GeV, discovery
of supersymmetry at LEPII or the Tevatron will not be possible
(except for the $\hl$ which
might be light enough to be observed if $\tanb$ is small). However,
observation of the gluino and squarks up to $\mgluino\sim 1-1.5$ TeV
should be straightforward at the LHC, and the $\cntwo,\cpmone,\slepr,\slepl$
(with masses of order 1/4 to 2/5 of $\mgluino$) can be easily discovered
and studied at an NLC with $\sqrt s$ in the 500 GeV to 1 TeV range.
Overall, the minimal-supergravity and dilaton-like
boundary conditions imply a very real
possibility of discovering supersymmetry at LEPII and/or the Tevatron,
and would certainly guarantee
exciting prospects for the future at LHC and an NLC.

\section{Other Recent Work}

The above establishes a basic phenomenological framework for discussing SUSY
detection. Of course, there were many
contributions to the conference relevant to this topic.  I give a brief
overview; the reader should consult the papers for details.

First, of course, are the steadily improving limits on the production of
SUSY particles at LEP.  These eliminate portions of the supersymmetric
parameter space, providing important constraints on model building.
Especially important are constraints deriving from non-observation of
$\chitil\chitil$ decays of the $Z$ \cite{lthree}, which provide critical
restrictions on the allowed domain in the $\mu$--$\mgluino$ parameter space
in the context of the MSSM.

Additional constraints that could be included in restricting the allowed
parameter space of a given model are those coming from $B-\anti B$ and
$K-\anti K$ mixing.  If $\mhpm$, $\mstopone$, and/or $\mcpmone$ are small and
$\tanb$ is small, then Ref.~\cite{bcko} shows that the associated
new loop-diagram contributions are such that consistency with the
well-established mixing results is not guaranteed, and such constraints
should be included.  Unfortunately, sensitivity to the additional loop diagrams
diminishes rapidly for $\tanb$ values above 2.

The impact of an invisibly decaying $\snu$ and
(possibly) $\cntwo$ upon SUSY detection (emphasized above),
was also examined for LEP200 and the Tevatron in Ref.~\cite{dgd}.

In models with $m^0>M^0$, or if we ignore the unification context altogether,
then one cannot entirely rule out the possibility that the gluino is very
light \cite{farrar}. As pointed out in Ref.~\cite{bb} (see
also references therein), a
light gluino can reconcile the apparent difference between $\alpha_s(\mz)$ as
extrapolated from deep inelastic scattering data (which yields a value of
about 0.108 in the absence of a light gluino), and the $\alpha_s(\mz)=0.122$
value extracted from LEP data.

Detection of the SUSY Higgs bosons is a critical issue.  Currently, there are
recent limits from LEP \cite{leplimits}. In the near future, we have seen
that Higgs detection
in the $Z\hl$ mode at LEPII can provide an important discovery channel for
SUSY. SUSY Higgs detection at hadron colliders is substantially more
difficult.
New modes for Higgs discovery at the Tevatron and LHC have been explored
during the course of the last year \cite{zeuthen}.
Techniques were developed for: a)
detection of an invisibly decaying $\hl$ in $W\hl$ and $t\anti t \hl$
associated production \cite{hinvisible}; and b)
using $b$-tagging to detect the $\hl$, $\hh$ and/or $\ha$ in their primary
$b\anti b$ decay modes via $W \,Higgs$, $t\anti t \,Higgs$
and $b\anti b \, Higgs$ associated production \cite{hbtagging}.
The results of these studies
are easily summarized. The most promising mode
at the Tevatron is $W\hl$ associated production
with $b$-tagging. However, for integrated luminosity of $10\fbi$ or less,
the Tevatron will at best probe $\mhl$ values up to the LEPII
discovery limits. In contrast, at the LHC the $b$-tagging
detection modes could ensure
that at least one of the SUSY Higgs bosons will be found,
in principle closing the famous hole
\cite{zeuthen} in parameter space where no SUSY Higgs would be found at the LHC
employing the standard $\gamma\gamma$ (or $\ell\gam\gam$) and $4\ell$ final
states. However, for these modes to be viable, it is necessary that
$b$-tagging be about 40\% efficient and 99\% pure (\ie\ no more than 1\%
mis-tagging probability for light quark and gluon jets). It is not yet clear
that the required efficiency and (especially) purity can be achieved in the
high luminosity, multiple-interaction LHC environment.

With regard to specific models, we note that the
supergravity/superstring models discussed earlier \cite{bgkp} all predict
relatively large masses (generally $\gsim 200\gev$) for the $\hh$, $\ha$ and
$\hpm$. The lighter $\hl$ is predicted to have mass below about
115 GeV  and relatively SM-like couplings. \Fref{dilfig} shows that
for much of parameter space,
the $\hl$ would be observable at LEPII with $\sqrt s =200$ GeV; and it
is {\it guaranteed} to be found at LEPII or an NLC with
$\sqrt s \gsim 230-250$ GeV. Observation of the $\hl$
at the LHC would also be possible. In contrast,
the heavier $\hh$, $\ha$ and $\hpm$ might escape observation.
At a hadron collider,
their decays to ino-pair and slepton-pair final states are important and
dilute normal detection modes, while the SUSY decay modes are not
easily employed in their own right.
The ability of a linear $\epem$ collider to observe $\ha\hh$ and $\hp\hm$
pair production (the only viable discovery modes when the
$\hh,\ha,\hpm$ are heavy) is
restricted by machine energy.  For $\sqrt s =500$ GeV,
discovery would only be possible for masses below about 220-230 GeV. Thus,
there is a significant chance
that these heavier MSSM Higgs bosons would also not be seen in the first
years of operation of the NLC (i.e. prior to upgrading the NLC
energy to $\sim 1\tev$).

Regarding the charged Higgs,
the CDF group has searched for $t\rta \hp b$
decays in $t\anti t$ events in dilepton final states.
In the context of a two-Higgs-doublet model
they exclude $\mhp\lsim \mt-\mb$ for $\mt\lsim 105\gev$
and large $\tanb$ values \cite{CDF}.  Of course, if $\mt\sim 170 \gev$
this analysis does not restrict $\mhp$.

In considering $\hp$ detection at either hadron or $\epem$ colliders,
one cannot ignore the possibility that $\hp\rta\wt t \,\wt b$ decays could
dominate if allowed \cite{hstop}, yielding much more complicated final state
signatures.

Extensions of the MSSM to include an additional singlet Higgs field
continue to be of interest.  In one contribution to this conference
\cite{NMSSM}, it is demonstrated that the lightest Higgs ($S_1$) in this
`NMSSM' model must have $m_{S_1}\leq 156\gev$. In related work, it was
shown that the $S_1$ or the next lightest $S_2$ is guaranteed to be
observable at an NLC (with $\sqrt s\geq 300\gev$),
via $ZS_{1,2}$ production, provided the
theory remains perturbative at all scales during RGE evolution \cite{NMSSMo}.

Of course, the supersymmetric extension of the SM may not turn out
to be that of the minimal model. If there is unifying group larger
than $SU(5)$, such as $E_6$ or a $L-R$ symmetric group, then one
must reassess the impact of the additional SUSY particles on the RGE
coupling unification and radiative electroweak symmetry breaking.  Assuming
that this program is successful, many new effects and signals for
SUSY could arise in such models. For example, in one contribution \cite{esix}
it is demonstrated that virtual effects from $E_6$ model interactions could
give rise to $\tau\rta eee$ and $Z\rta e\tau$ decays at an observable level.
In $L-R$ models \cite{LR}, the doubly-charged Higgsino ($\wt \Delta^{++}$)
not only enhances slepton pair production in $\epem$ collisions via virtual
$u$-channel exchange diagrams, but also would provide a number of unusual and
clean signals when produced directly in $\epem$, $e^-e^-$, $e^-\gamma$, and
$\gamma\gamma$ collisions.

If $R$-parity violation is present in the supersymmetric theory, then
detection phenomenology undergoes considerable change.  Early
work \cite{Rparity} pointing out the importance of leptonic signatures (as
opposed to missing energy signatures) has recently been extended in
Ref.~\cite{spectacular} to show that these leptonic signatures could be
spectacular if the {\it only} source of $R$-parity violation is
a superpotential term of the type $W\ni L_iL_jE_k$, where the $L$
and $E$ superfields are those for the lepton doublets and singlets,
respectively.  In this case, the LSP $\cnone$ decays entirely to visible
leptons, and the importance of leptonic signatures is apparent.

Finally I mention the results of Ref.~\cite{fourthgen} in which the
possibility of including a fourth generation in the standard MSSM context is
considered.  The result is that this remains a possibility
without violating perturbative limits on {\it any} of the Yukawa couplings,
including those associated with the fourth generation, but only if
exact $b-\tau$ Yukawa unification is not required.  The allowed $t'$ and $b'$
masses will be accessible with increased luminosity at the Tevatron.
For $\mt\gsim 160\gev$, the
$\nu'$ and $\tau'$ masses are such that these new leptons would certainly be
observed at LEPII.

\section{Conclusions}

Overall, we see that the Tevatron with $L=1\fbi$ and LEPII with
$\sqrt s=200\gev$ and $L=500\pbi$ would be relatively complementary
in searching for SUSY in the minimal-supergravity and dilaton-like
superstring/supergravity-motivated MSSM
models outlined earlier. Experimentalists should be encouraged by the
relatively large values of the $\mgluino$ parameter that can be explored by
combining these two machines.  If these kinds of models are correct, we may
not have to wait until the LHC and/or an NLC to find the first signal for SUSY.

New modes for Higgs detection at hadron colliders have been developed,
and, at the LHC,  show
considerable promise for providing a true guarantee that at least one
of the MSSM Higgs bosons will be detectable.  In this regard,
$b$-tagging at the LHC could prove crucial,
and deserves maximal effort on the part
of the LHC detector collaborations.

Finally, supersymmetric models with extended gauge groups, $R$-parity
violation,
or a fourth generation all provide
interesting new phenomena and experimental signatures, many of which could
prove to be of particular interest at Tevatron energies.


\end{document}